\documentstyle[prd,aps,floats,amsfonts,amssymb,amsmath]{revtex}

\begin{document}
 \draft
  \title{Gauge Invariant Wave Equations in Curved Space-Times and Primordial Magnetic Fields}
\author{G. Lambiase$^{a,b}$ and A.R. Prasanna$^c$}
\address{$^a$Dipartimento di Fisica "E.R. Caianiello", Universit\'a di Salerno, 84081 Baronissi (Sa), Italy.}
  \address{$^b$INFN - Gruppo Collegato di Salerno, Italy.}
\address{$^c$Physical Research Laboratory, Navrangpura, Ahmedabad 380 009, India.}
\date{\today}
\maketitle
\begin{abstract}
The inflationary production of magnetic field seeds for galaxies
is discussed. The analysis is carried out by writing the wave
equation of the electromagnetic field in curved spacetimes. The
conformal invariance is broken by taking into account of the
interaction of the electromagnetic field with the curvature tensor
of the form $\lambda
R_{\alpha\beta\gamma\delta}F^{\alpha\beta}F^{\gamma\delta}$. Such
a term induces an amplification of the magnetic field during the
reheating phase of the universe, but no growth of the magnetic
field occurs in the de Sitter epoch. The resulting primordial
magnetic field turns out to have strengths of astrophysical
interest.
\end{abstract}
% \Keyword(s){Particle theory and field theory models
%of the early universe, Electric and Magnetic Field}
%\vskip2pc]

%\section{Introduction}
%\setcounter{equation}{0}

One of the open issue of modern astrophysics is the origin of
intragalactic magnetic fields, with the characteristics that their
amplitudes are of the order of $\sim 10^{-6}$G and they are
uniform on scales of the order of $\sim 10$kpc.

A promising candidate to explain the primordial magnetic field
generation is the dynamo mechanism discussed by Zeldovich,
Ruzmaikin and Sokoloff \cite{zeldovich}, and by Parker
\cite{Parker}. The dynamo effect induces an amplification of a
preexisting magnetic field, but it requires a seed field at the
epoch of galaxy formation coherent over a scale of $\sim 1$Mpc.
The question that arises is about the mechanism that produced such
a seed field.

Starting with the observation that the Universe, during its
evolution, has behaved as a good conductor, one expects that the
evolution of a primordial magnetic field may preserve the magnetic
flux. This physical aspect is encoded in the parameter
$r=\rho_B/\rho_\gamma$, which remains (with good approximation)
constants and provides an invariant measure of magnetic field
strength. Here $\rho_B=|{\bf B}|^2/8\pi$ is the energy density of
a magnetic field, and $\rho_\gamma=\pi^2T^4/25$ is the energy
density of the cosmic microwave background radiation. In order to
explain the present value of $r\approx 1$ for galaxies, one needs
a pre-galactic magnetic field to which corresponds (see for
example, Refs. \cite{turner,mazzitelli})
 \begin{equation}\label{0}
 r\simeq 10^{-34}\,.
\end{equation}
Turner and Widrow have suggested that a magnetic field might be
generated by quantum fluctuations during an inflationary epoch,
and it could be sustained after the wave length of interest
crossed beyond the horizon giving the observed field today
\cite{turner}. This model invokes a coupling among the
electromagnetic field and the curvature tensors $R$ and
$R_{\alpha\beta}$, which break the conformal invariance.
Nevertheless the field equations derived in Ref. \cite{turner}
{\it are not} gauge invariant. Turner-Widrow's model has been
generalized by Garretson, Field, and Carroll \cite{carroll} by
introducing a coupling of the photon with an arbitrary pseudo
Goldstone boson (see also \cite{jackiw}). Ratra introduced a
coupling between the scalar field $\phi$ and the electromagnetic
field via the interaction $e^\phi F_{\mu\nu}F^{\mu\nu}$, where
$F_{\mu\nu}$ is the electromagnetic field strength,
$F_{\mu\nu}=\nabla_\mu A_\nu-\nabla_\nu A_\mu$. \cite{ratra}. The
generation of primordial magnetic field induced by coupling terms
of the form $R^nF_{\mu\nu}F^{\mu\nu}$, which preserves the gauge
invariance, has been studied by Mazzitelli and Spedalieri
\cite{mazzitelli}, whereas Dolgov has studied the breaking of
conformal invariance in QED due to trace anomalies \cite{dolgov}.
For other mechanisms aimed to explain in what way the primordial
magnetic field might be generated, see Refs.
\cite{dimopoulos,vashaspati,kibble,opher,gasperini,bertolami,calzetta,maroto,veneziano,enqvist,widrowRMP,grasso}.

This paper concerns the origin of the primordial magnetic field in
the framework of the non-minimal coupling of electromagnetic field
in curved space time, whose action is given by
\cite{prasannaCQG,prasannaPLA}
 \begin{equation}\label{action}
S= \int d^4x\sqrt{-g}\left(
    -\frac{1}{4} F_{\mu\nu}F^{\mu\nu}
    +\lambda R^{\mu\nu\rho\sigma}F_{\mu\nu}F_{\rho\sigma}
    \right)\,.
 \end{equation}
where the $\lambda$-term, although invariant under CPT and general
coordinate transformations, violates the Einstein's equivalence
principle. The motivation to introduce the nonminimal coupling of
electromagnetic field with the gravitational field
\cite{prasannaPLA} was essentially to understand the intricate
relation between curvature and all forms of energy and their
variations. Though at the present epoch, this coupling could be
very small for cosmological background ($\lambda<
10^{11}$cm$^2\sim 2.5\times 10^{20}$eV$^{-2}$ \cite{prasannaCQG}),
at the very early universe the effect of the nonminimal terms
could become important, apart from breaking the conformal
invariance which indeed is necessary for the generation of large
enough magnetic seed from the vacuum. The fact that this
lagrangian preserves the gauge invariance but breaks the conformal
invariance and gets very small at the present epoch is indeed a
very satisfactory situation. In our approach of writing the wave
equation for the fields \cite{prasannaNPB}, we get additional
contributions from the curvature terms apart from the nonminimal
terms that have been shown to arise from one loop quantum
corrections \cite{Drummond}, indicating the qualitative
significance of such a coupling in the early universe scenario.

Variation with respect to the four-potential $A_\mu$ yields the
field equations \cite{prasannaCQG}
\begin{equation}\label{fieldequation}
  \nabla_\alpha F^{\alpha\mu}=-2\lambda \left[R^{\mu\nu\rho\sigma}\nabla_\nu F_{\rho\sigma}
  +(\nabla_\rho R^\mu_{\phantom{\mu}\sigma})F^{\rho\sigma}\right]\,.
\end{equation}
We also note that the electromagnetic field $F_{\mu\nu}$ satisfies
the Bianchi identities
\begin{equation}\label{bianchi}
  \nabla_\mu F_{\nu\lambda}+\nabla_\nu F_{\lambda\mu}+\nabla_\lambda
  F_{\mu\nu}=0\,.
\end{equation}
Eqs. (\ref{fieldequation}) and (\ref{bianchi}) give the Maxwell
equations in curved spacetimes. To infer the wave equations for
$F_{\mu\nu}$, we apply the covariant derivative $\nabla_\lambda$
to Eq. (\ref{fieldequation}), and by making use of Eq.
(\ref{bianchi}) we get
\begin{equation}\label{5}
  \Box F_{\nu\lambda}+[\nabla^\mu,
  \nabla_\nu]F_{\lambda\mu}-[\nabla^\mu, \nabla_\lambda]F_{\nu\mu}=-2\lambda
  \left(
  \nabla_\lambda[R^{\mu\nu\rho\sigma}(\nabla_\nu F_{\rho\sigma})]+
  \nabla_\lambda[(\nabla^\rho R_\lambda^{\phantom{\lambda}\sigma})F_{\rho\sigma}]
  \right)\,,
\end{equation}
where $\Box =\nabla^\mu\nabla_\mu$ and $[.,.]$ is the commutator.
The Ricci identity \cite{carmeli}
\begin{equation}\label{6}
  [\nabla^\mu, \nabla_\nu]F_{\alpha\mu}=
  R_{\rho\alpha\nu\mu}F^{\rho\mu}+R^\rho_{\phantom{\rho}\nu} F_{\alpha\rho}\,,
\end{equation}
and the well known cyclic identities of Riemann's tensor
\begin{equation}\label{7}
  R_{\rho\alpha\beta\gamma}+R_{\rho\gamma\alpha\beta}+R_{\rho\beta\gamma\alpha}=0\,,
\end{equation}
allow to cast Eq. (\ref{5}) in the form
\begin{equation}\label{eqGeugeInv0}
  \Box F_{\nu\lambda}+R^{\rho\mu}_{\phantom{\rho\mu}\nu\lambda}F_{\rho\mu}+
  R^\rho_{\phantom{\rho}\nu} F_{\lambda\rho}- R^\rho_{\phantom{\rho}\lambda} F_{\nu\rho}=
  2\lambda\left(
  \nabla_\nu[R^{\alpha\phantom{\lambda}\rho\sigma}_{\phantom{\alpha}\lambda}(\nabla_\alpha
  F_{\rho\sigma})]+
  \nabla_\nu [(\nabla^\rho
  R_{\lambda}^{\phantom{\lambda}\sigma})F_{\rho\sigma}]-(\nu\leftrightarrow\lambda)
  \right)\,.
\end{equation}
We now observe that the necessary and sufficient condition to have
a system of coordinates in which the metric is conformal to the
Minkowski one is that the Weyl tensor $C_{\lambda\mu\nu\rho}$
vanishes. As a consequence, the Riemann tensor can be written in
terms of the Ricci tensor and the scalar curvature $R$
\cite{weinberg}, i.e. (in $N$ dimensions)
 \[
 R_{\lambda\mu\nu\rho}=\frac{1}{N-2}(g_{\lambda\nu}R_{\mu\rho}-g_{\lambda\rho}R_{\mu\nu}
 -g_{\mu\nu}R_{\lambda\rho}+g_{\mu\rho}R_{\lambda\nu})-\frac{R}{(N-1)(N-2)}(g_{\lambda\nu}g_{\mu\rho}-
 g_{\lambda\rho}g_{\mu\nu})\,.
 \]
After some algebras, for conformally flat spacetimes ($N=4$) Eq.
(\ref{eqGeugeInv0}) simplifies to the form
\begin{eqnarray}\label{eqGaugeInv}
  \Box F_{\nu\lambda}-\frac{R}{3}F_{\nu\lambda}&=&
  2\lambda\left(
    \nabla_\nu[R^{\alpha\phantom{\lambda}\rho\sigma}_{\phantom{\alpha}\lambda}(\nabla_\alpha
  F_{\rho\sigma})]+
  \nabla_\nu [(\nabla^\rho
  R_{\lambda}^{\phantom{\lambda}\sigma})F_{\rho\sigma}]-(\nu\leftrightarrow\lambda)
  \right) \\
  &=& 2\lambda \left[\nabla_\nu \left(R_{\lambda\sigma}\nabla_\rho F^{\rho\sigma}-
        R_{\alpha\sigma}\nabla^\alpha
        F_{\lambda}^{\phantom{\lambda}\sigma}-\frac{R}{3}\nabla^\alpha
        F_{\alpha\lambda}\right)+\nabla_\nu[(\nabla^\rho R_\lambda^{\phantom{\lambda}\sigma})F_{\rho\sigma})]-
        (\lambda \leftrightarrow \nu)\right] \nonumber
        \,.
\end{eqnarray}
The metric components of the spatially flat
Friedman-Robertson-Walker cosmology are, in the conformal time,
\begin{equation}\label{conf-metric}
  g_{\mu\nu}=a^2(\eta) \, \mbox{diag} (1, -1, -1, -1)\,,
\end{equation}
where $a(\eta)$ is the scale factor. The field strength tensor
$F_{\mu\nu}$ has components
\begin{equation}\label{Fmunu}
  F_{\mu\nu}=a^2(\eta) \begin{pmatrix}
                       0  & -E_x & -E_y & -E_z  \\
                       E_x & 0 & B_z & -B_y \\
                       E_y & -B_z & 0 & B_x \\
                       E_z & B_y & -B_x & 0 \\
                       \end{pmatrix} \,.
\end{equation}
whereas Riemann's and Ricci's tensor components are
\begin{equation}\label{RiemComponents}
  R^{ij}_{\phantom{ij}kl}=-\frac{\dot{a}^2}{a^4}(\delta^i_k\delta^j_l-\delta^i_l\delta^j_k)\,,\quad
  R^{0i}_{\phantom{0i}0j}=-\frac{1}{a^2}\left(\frac{\ddot{a}}{a}-\frac{\dot{a}^2}{a^2}\right)\delta^i_j\,,
  \quad  R^{ij}_{\phantom{ij}0k}=0\,,
\end{equation}
\begin{equation}\label{RComponents}
  R^i_{\phantom{i}j}=-\frac{1}{a^2}\left(\frac{\ddot{a}}{a}+\frac{\dot{a}^2}{a^2}\right)\delta^i_j
  \equiv - u(\eta)\delta^i_j\,,
  \quad R^0_{\phantom{0}0}=-\frac{3}{a^2}\left(\frac{\ddot{a}}{a}-\frac{\dot{a}^2}{a^2}\right)\equiv - s(\eta)\,,
  \quad R^i_{\phantom{0}0}=0\,,
\end{equation}
where the dot stands for derivative with respect to the conformal
time $\eta$. Observing that in the conformal metric
(\ref{conf-metric}), the LHS of Eq. (\ref{eqGaugeInv}) for the
spatial component assumes the form
\begin{equation}\label{LHS}
 \Box F_{ij}-\frac{R}{3}F_{ij}=\frac{1}{a^2}\, \Box_\eta F_{ij}\,,
\end{equation}
where $\Box_\eta$ is the D'Alambertian in the Minkowski spacetime,
i.e.
$\Box_\eta=\eta^{\mu\nu}\partial_\mu\partial_\nu=\partial^2/\partial
\eta^2-\nabla_\eta^2$, it follows that Eq. (\ref{eqGaugeInv})
becomes
\begin{equation}\label{RHS}
 \Box_\eta F_{ij} = 2\lambda\left\{
 -\frac{2s}{3}\, \ddot{F}_{ij}+
 \left(u-\frac{s}{3}\right)\nabla^2_\eta F_{ij}+
 \left[5(s-u)\,\frac{\dot{a}}{a}-\dot{u}\right]\,\dot{F}_{ij}-
 6(s-u)\left(\frac{\dot{a}}{a}\right)^2F_{ij}\right\}\,,
\end{equation}
where $s$ and $u$ have been defined in (\ref{RComponents}). In
terms of the magnetic field and collecting the terms, Eq.
(\ref{RHS}) reads
\begin{equation}\label{EqWaveExplicit}
  \left(1+\frac{4\lambda}{3}\,s\right)\,\frac{\partial^2}{\partial \eta^2}(a^2{\bf B})-
  \left[1+2\lambda\left(u-\frac{s}{3}\right)\right]\nabla^2_\eta (a^2{\bf B})-
  2\lambda\left[5(s-u)\frac{\dot{a}}{a}-\dot{u}\right]\frac{\partial}{\partial \eta}(a^2{\bf B})+
  12\lambda(s-u)\left(\frac{\dot{a}}{a}\right)^2(a^2{\bf B})=0\,.
\end{equation}
Taking the spatial Fourier transform
\begin{equation}\label{FourierTransf}
  {\bf B}(\eta, {\bf k})=\int \frac{d^3{\bf x}}{2\pi}\, e^{i{\bf k}\cdot {\bf
  x}}\,{\bf B}(\eta, {\bf x})\,,
\end{equation}
and using the notation
\begin{equation}\label{FNOtation}
  {\bf F}_k(\eta)=a^2(\eta){\bf B}(\eta, {\bf k})\,,
\end{equation}
Eq. (\ref{EqWaveExplicit}) can be rewritten as
\begin{equation}\label{EqWaveF}
  \left(1+\frac{4\lambda}{3}\,s\right)\ddot{{\bf F}}_k
 -2\lambda\left[5(s-u)\frac{\dot{a}}{a}-\dot{u}\right]\dot{{\bf F}}_k
 +\frac{n(\eta)}{\eta^2}{\bf F}_k=0 \,,
\end{equation}
where
\begin{equation}\label{n(eta)}
  n(\eta)=\eta^2\left\{\left[1+2\lambda\left(u-\frac{s}{3}\right)\right]k^2
  +12\lambda (s-u)\left(\frac{{\dot a}}{a}\right)^2\right\}\,.
\end{equation}
Let us now evaluate the magnetic field for the different phases of
evolution of the Universe \cite{turner}. In what follows we
concern with the evolutions of the magnetic field fluctuations
whose wavelengths are well outside the horizon, according to
\cite{turner}, $L_{phys}=aL\gg H^{-1}$ or $k\eta\ll 1$. We shall
use the notation $F_k=|{\bf F}_k|=\sqrt{{\bf F}^*_k\cdot {\bf
F}_k}$.

\vspace{0.2in}

\noindent {\it Inflationary de Sitter (dS) phase}. The scale
factor for this epoch of the Universe is
\begin{equation}\label{ScaleFactorInfl}
  a(\eta)\sim -\frac{1}{H_{dS}\eta}\,,
\end{equation}
where $H_{dS}\sim 3\times 10^{24}$eV \cite{bertolami}. Eq.
(\ref{EqWaveF}) reduces to $(1+4\lambda H_{dS}^2)({\ddot
F}_k+k^2F_k)=0$, whose solution is $F_k(\eta) \sim \sin k\eta$. It
is worth to note that the $\lambda$-term does not affect the
growth of the magnetic field. For modes outside the horizon, we
have in fact that $F_k$ depends on $\eta$ as
\begin{equation}\label{F(a)Infl}
  F_k(\eta) \sim \eta \sim a^{-1}\,.
\end{equation}

\vspace{0.2in}

\noindent {\it Phase of Reheating (RH) and Matter Domination}. The
scale factor for this stage of the Universe is
\begin{equation}\label{ScaleFactRH}
  a(\eta)\sim \frac{1}{4}H_0^2R_0^3\eta^2\,,
\end{equation}
where $R_0\sim 10^{26}h_0^{-1}$m ($0.6\leq h_0\leq 0.8$) is the
present Hubble radius of the Universe, and $H_0\sim 100 h_0
$km/Mpc sec is the Hubble parameter. The expressions for $u$ and
$s$, defined in (\ref{RComponents}),
 \[
 u(\eta)=\frac{6}{b^2}\, \frac{1}{\eta^6}\,, \qquad
 s(\eta)=-u(\eta)\,.
 \]
where $b\equiv 16\lambda/(H_0^2R_0^3)^2$, and the condition that
the $\lambda$-term is dominant $\eta\ll b^{1/6}$, allow to rewrite
Eq. (\ref{EqWaveF}) in the form
\begin{equation}\label{EqWaveFRH}
  \ddot{\bf F}_k
  - \frac{21}{\eta}\,\dot{\bf F}_k
  + \frac{72}{\eta^2}{\bf F}_k=0\,,
\end{equation}
whose solution is
\begin{equation}\label{F(a)RH}
  F_k(\eta) \sim \eta^{18}\sim a^9\,.
\end{equation}
Eq. (\ref{F(a)RH}) shows that the magnetic field grows at very
high power of scale factor $a$, which is the main amplification
for the seed of primordial magnetic fields.

\vspace{0.2in}

\noindent {\it Phase of Radiation Domination (RD)}. The scale
factor of the Universe is
\begin{equation}\label{ScaleFactRD}
  a\sim H_0R_0^2\eta\,.
\end{equation}
Eq. (\ref{EqWaveF}) reduces
\begin{equation}\label{EqWaveFRD}
  \left(1-\frac{4d}{\eta^4}\right)\ddot{\bf F}_k-
  \frac{32d}{\eta^5}\,\dot{\bf F}_k-\frac{48 d}{\eta^6}{\bf
  F}_k=0\,,
\end{equation}
where $d\equiv \lambda/(H_0R_0^2)^2$. The solution is
\begin{equation}\label{SolutionRD}
 F_k(\eta)=c_1'\, \frac{12 d+5\eta^4}{12d\eta^4}
 +c_2'\frac{20d+3\eta^4}{20d\eta^3}\,.
\end{equation}
Using as boundary condition $F_k \to constant$ for increasing
conformal time, we get $c_2'=0$. Hence the solution is
$F_k(\eta)\sim 5/12d+\eta^{-4}$. The last term is suppressed
during the evolution of the universe thus $F_k \sim constant$,
which implies $\rho_B\sim a^{-4}$, as expected.

\vspace{0.1in}

The expressions of $F_k(\eta)$ for different epochs of the
evolution of the Universe allow to estimate the strength of the
primordial magnetic field. According to Turner-Widrow's model
\cite{turner}, if one assumes that the Universe had gone through a
period of inflation at GUT scale ($M_{GUT}\sim 10^{16}\div
10^{17}$GeV) and that fluctuations of the electromagnetic field
have come out from the horizon where the Universe had gone through
about 55 e-folding of inflation, then \cite{turner}
\begin{equation}\label{r}
  r\approx (7\times
  10^{25})^{-2(p+2)}\left(\frac{M_{GUT}}{m_{Pl}}\right)^{4(q-p)/3}
  \left(\frac{T_{RH}}{m_{Pl}}\right)^{2(2q-p)/3}
  \left(\frac{T_{*}}{m_{Pl}}\right)^{-8q/3}L^{-2(p+2)}_{Mpc}\,,
\end{equation}
where $T_{RH}$ is the reheating temperature, $T_*$ is the
temperature at which plasma effects become dominant (i.e. the
Universe first becomes a good conductor), and $m_{Pl}\sim
10^{19}$GeV is the Planck mass. Finally, $p$ and $q$ are the
exponents of the scale factor $a(\eta)$ during the dS and RH
epochs which determine an increasing of the magnetic field. In our
case $p=-1$ and $q=9$ (see Eqs. (\ref{F(a)Infl}) and
(\ref{F(a)RH})). The temperature $T_*$ can be estimated via
reheating processes \cite{turner} $T_*=min\{(T_{RH}M_{GUT})^{1/2};
(T^2_{RH}m_{Pl})^{1/3}\}$, and for $T<T_*$ $\rho_B$ evolves as
$\rho_B\sim a^{-4}$. We note, however, that the reheating
temperature $T_{RH}$ is given by $T_{RH}=\{10^{9}\mbox{GeV},
M_{GUT}\}$. In Table I are reported the values of $r$. The range
of variability of $r$ is $ r\sim  10^{-50}\div 10^{-32}$, and we
also note that because of the contribution of $\lambda$ the value
for the seed field becomes extremely sensitive to the values of
the physical parameters entering in $r$. In particular,
$M_{GUT}\sim 10^{17}$GeV, $T_{RH}\sim 10^{16}$GeV, and $T_*\sim
10^{15.56}$GeV imply $r\sim 10^{-34}$.

In conclusion, the non-minimal coupling of the electromagnetic
field with curvature terms, which breaks the conformal invariance
but preserves the gauge invariance of the field wave equations,
provides a scenario in which magnetic fields may be amplified.
Such an growing occurs mainly during the reheating phase, whereas
no amplification arises in the inflationary stage. This picture
allows to infer the order of magnitude of $r$ necessary to
generate the seed of galactic magnetic fields.

\begin{table}
\caption{Values of $r=\rho_B/\rho_\gamma$ at 1Mpc and for
$M_{GUT}\sim 10^{16}\div 10^{17}$GeV, $T_{RH}=\{10^9, 10^{15}-
10^{17}\}$GeV, and $T_*=\{10^{12}, 10^{15}, 10^{16}\}$GeV [3].}
\begin{tabular}{llll}
 $M_{GUT}$(GeV) & $T_{RH}(GeV)$ & $T_*$(GeV) & $r\sim$  \vspace{0.09in} \\
  $10^{17}$ & $10^9$ & $10^{12}$ &  $10^{-37}$ \\
             & $10^{15}$ & $10^{15}$ &  $10^{-33}$ \\
             & $10^{16}$ & $10^{16}$ &  $10^{-45}$ \\
             & $10^{16}$ & $10^{15.56}$ & $10^{-34}$ \\
             & $10^{17}$ & $10^{16}$ & $10^{-32}$ \vspace{0.09in} \\
  $10^{16}$ & $10^{9}$ & $10^{12}$ &  $10^{-50}$ \\
\end{tabular}
\end{table}
%\vspace{0.5in}

%\acknowledgments
%The authors thank the referees for comments which improved the
%paper. Research supported by PRIN 2003.

\end{document}